\documentclass[twocolumn]{webofc_arxiv}
\usepackage[varg]{txfonts}   

\usepackage{stfloats}
\usepackage{amsmath,amssymb}
\usepackage[bookmarks, colorlinks=true, linktoc=page, pdftex, linkcolor=red, citecolor=red, urlcolor=red]{hyperref}

\newcommand{\conex}{\textsc{conex}}
\newcommand{\corsika}{\textsc{corsika}}

\newcommand{\pythia}{\textsc{pythia}}
\newcommand{\epos}{\textsc{epos}}
\newcommand{\eposlhc}{\textsc{epos-lhc}}
\newcommand{\qgsjet}{\textsc{qgsjet}} 
\newcommand{\qgsjetII}{\textsc{qgsjet-ii-04}} 
\newcommand{\sibyll}{\textsc{sibyll}}

\newcommand{\sqrts}{\sqrt{s}}
\newcommand{\sqrtsnn}{\sqrt{s_{_{\rm NN}}}}

\newcommand{\pT}{p_{_{\rm T}}}
\newcommand{\Qsat}{Q_{\rm sat}}
\newcommand{\ECR}{E_{_{\rm CR}}}
\newcommand{\Xmax}{\left<X_{\rm max}\right>}

\newcommand{\smax}{\sigma_{X_{\rm max}}}

\newcommand{\meanpt}{\rm \left< \pT \right>}
\newcommand{\dNdeta}{dN_{\rm ch}/d\eta|_{\eta=0}}

\newcommand*{\eg}{e.g.,}

\newcommand*{\cm}{c.m.}

\newcommand*{\elm}{e.m.}

\begin{document}
\setcounter{page}{1}
%


\title{Ultrahigh-energy cosmic rays: Anomalies, QCD, and LHC data}

\author{ David d'Enterria\inst{1}\fnsep\thanks{\email{david.d'enterria@cern.ch}}}
\institute{CERN, EP Department, 1211 Geneva, Switzerland}
\abstract{
Measurements of proton and nuclear 
collisions at the Large Hadron Collider at nucleon-nucleon \cm\ energies up to $\sqrtsnn=$~13~TeV
have improved our understanding of hadronic interactions at the highest energies 
reached in collisions of cosmic rays with nuclei in the earth atmosphere, up to
$\sqrtsnn\approx 450$~TeV. The Monte Carlo event generators (\epos, \qgsjet, and \sibyll) 
commonly used to describe the air showers generated by ultrahigh-energy cosmic rays 
(UHECR, with $\ECR\approx 10^{17}$--$10^{20}$~eV) feature now, after parameter retuning 
based on LHC Run-I data, more consistent predictions on the 
nature of the cosmic rays at the tail of the measured spectrum. 
However, anomalies persist in the data that cannot be accommodated by the models. 
Among others, the total number of muons 
(as well as their maximum production depth) remains significantly underestimated (overestimated) 
by all models. Comparisons of \epos, \qgsjet, and \sibyll\ predictions 
to the latest LHC data, and to collider MC generators such as \pythia, indicate that improved description 
of hard multiple minijet production and nuclear effects 
may help reduce part of the data--model discrepancies, shed light on the UHECR composition approaching 
the observed $\ECR\approx 10^{20}$~eV cutoff, and uncover any potential new physics responsible 
for the observed anomalies.
}
\maketitle
%
\section{Introduction}
\label{intro}

Ultrahigh-energy cosmic rays (UHECR), with energies $\ECR\approx 10^{17}$--$10^{20}$~eV, are produced
and accelerated in (poorly-known) extreme astrophysical environments. 
Their exact extragalactic sources and their nature, protons or heavier ions, remain still open questions 
today~\cite{Mollerach:2017idb,Dawson:2017rsp}. When reaching earth, they collide with N,\!\,O nuclei 
in the upper atmosphere at \cm\ energies, $\sqrts = \sqrt{\rm \,m_{_{\rm CR}}^2+m_{_{\rm N,O}}^2+2\cdot\ECR\cdot m_{_{\rm CR}}}
\approx \sqrt{\rm \,2\cdot10^{9}\cdot \ECR(\rm eV)}\approx14$--450~TeV, up to 30 times larger than those ever reached 
in any human-made collider~\cite{Bluemer:2009zf}.  UHECR produce gigantic cascades of secondary particles in the atmosphere, 
called extensive air showers (EAS)~\cite{Kampert:2012mx}, measured in dedicated observatories, such as the Pierre 
Auger Observatory~\cite{ThePierreAuger:2015rma} and Telescope Array (TA)~\cite{AbuZayyad:2012ru}, that combine 
the information from (i) the lateral distributions of secondary particles in surface detectors, and (ii) 
the fluorescence photons produced by nitrogen molecules excited along the shower. Key EAS observables are the 
average depth of the shower maximum $\Xmax$ and the RMS width of its fluctuations $\smax$, and the number and total 
energy of electrons ($N_{\rm e}, E_{\rm e}$) and muons ($N_{\rm \mu}, E_{\rm \mu}$) on the ground for various 
shower zenith angle ($\theta$) inclinations. By comparing these shower properties to the predictions of Monte Carlo 
(MC) models, the energy and mass of the incoming particles can be inferred.

\begin{figure*}[htp!]
\centering
\includegraphics[width=0.58\textwidth]{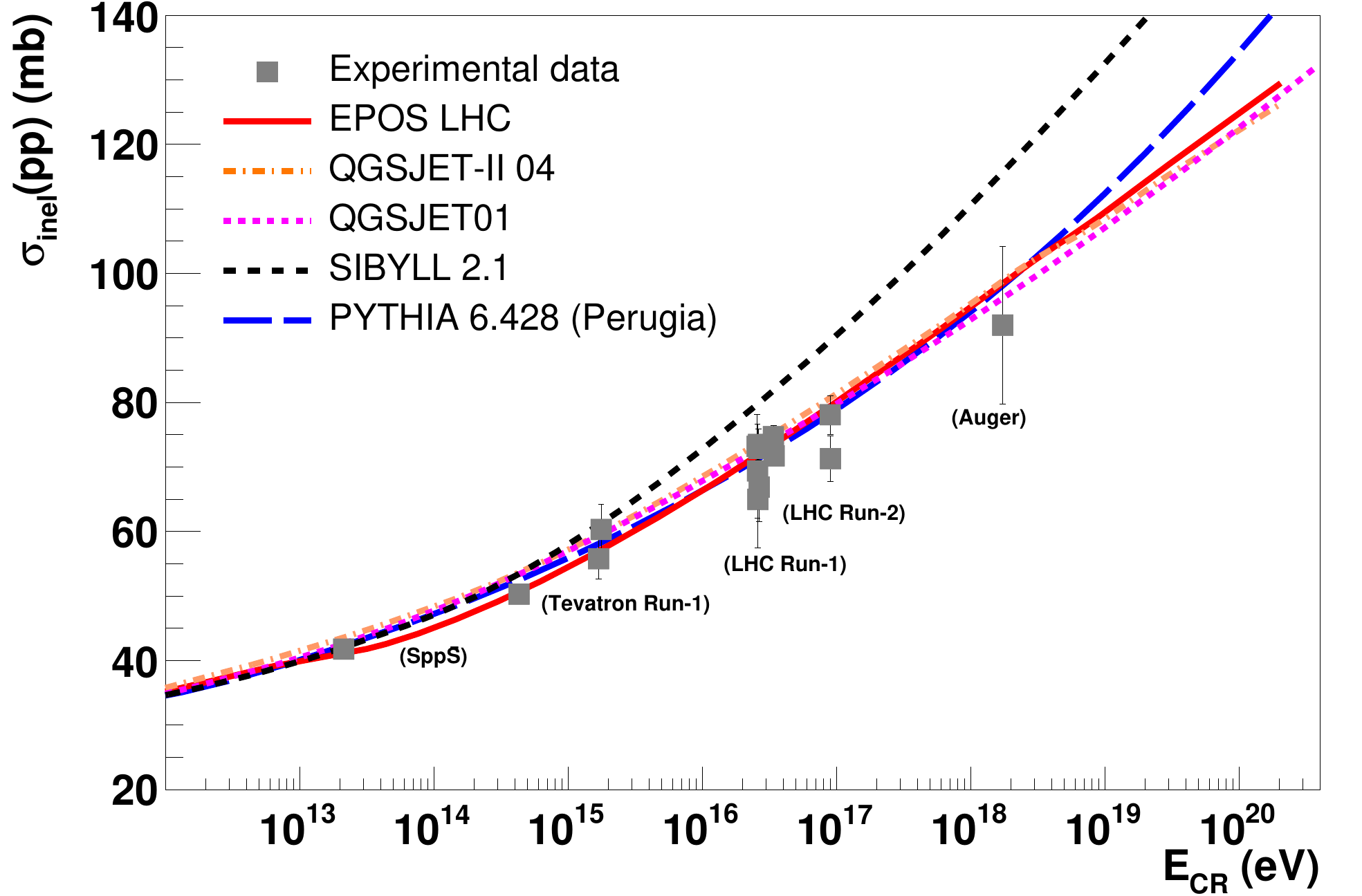}
\includegraphics[width=0.40\textwidth]{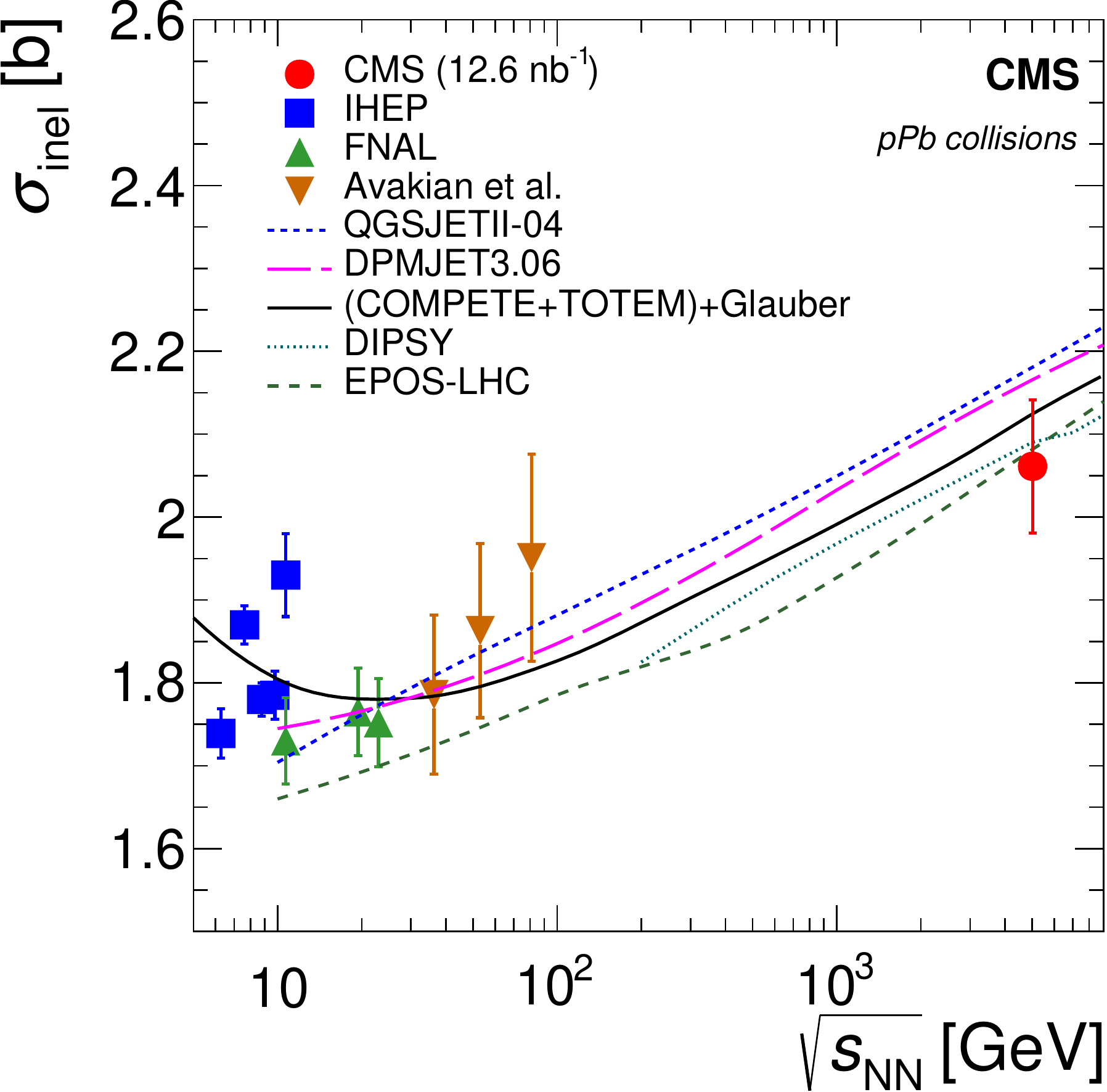}
\caption{Left: Inelastic p-p cross section as a function of CR energy: Data compiled in~\cite{dEnterria:2011twh,dEnterria:2016oxo}
compared to MC predictions retuned (except for \sibyll) to LHC Run-1 results~\cite{dEnterria:2018kcz}. 
Right: Inelastic p-Pb hadronic cross section (data and MC predictions) as a function of nucleon-nucleon 
collision energy~\cite{Khachatryan:2015zaa}.}
\label{fig:sigma_vs_sqrts}
\end{figure*}

The comparison of the measured EAS properties to the MC predictions is commonly done by interfacing the
\corsika~\cite{Heck:1998vt} or \conex~\cite{Bergmann:2006yz} air transport programs to event generators such as \eposlhc~\cite{Pierog:2013ria}, 
\qgsjet~01~\cite{Kalmykov:1993qe}, \qgsjetII~\cite{Ostapchenko:2010vb}, \sibyll~2.1~\cite{Ahn:2009wx}, 
or \sibyll~2.3c~\cite{Riehn:2017mfm} for the hadronic interactions, plus {\sc egs}4~\cite{Nelson:1985ec} for the 
electromagnetic (\elm) cascade evolution. 
The hadronic interaction models describe particle production in high-energy proton and nuclear collisions 
based on 
the Gribov's Reggeon Field Theory (RFT) framework~\cite{Gribov:1968fc}, extended 
to take into account perturbative quantum chromodynamics (pQCD) scatterings in (multiple) hard
parton collisions via ``cut (hard) Pomerons'' (identified diagrammatically as a ladder of gluons). 
With parameters tuned to reproduce the existing accelerator data~\cite{dEnterria:2011twh}, 
all those MC generators are able to describe the overall EAS properties, although anomalies persist in the 
UHECR results that cannot be accommodated by the models. On the one hand, the $\Xmax$ and $\smax$ dependence on $\ECR$ indicates 
a change of cosmic-ray composition from proton-dominated to a heavier mix above $\ECR \approx 10^{18.5}$~eV~\cite{Aab:2014kda,Abbasi:2014sfa},
but quantitative differences among the $\Xmax$ and $\smax$ predictions exist 
that are not fully understood, even though independently each MC generator globally reproduces the LHC 
data~\cite{Ostapchenko:2014mna,Pierog:2015epa,Ostapchenko:2016wtv}. 
On the other hand, in the same range of $\ECR$ energies, many measurements~\cite{Dembinski18} 
indicate significantly larger muon yields on the ground in particular at large transverse distances from the shower 
axis~\cite{AbuZayyad:1999xa,Aab:2014pza}, as well as lower $\mu$ production depths~\cite{Aab:2014dua}, 
than predicted by all models.

The difficulties in reconciling aspects of the UHECR data with the MC generator predictions suggest that the best models of hadronic 
interactions are missing some physics ingredient. For the muon component, either they do not account for processes that produce harder muons,
such as from, \eg\ jets or heavy-quark (in particular charm~\cite{Engel:2015dxa}) decays, and/or they do not feed enough energy into 
the EAS hadronic component (such as, \eg\ through an increased production of baryons~\cite{Pierog:2006qv}). 
More speculative explanations have been suggested based on changes in the physics of the strong interaction at energies beyond 
those tested at the LHC~\cite{Allen:2013hfa,Aloisio:2017ooo}, or on the production of electroweak sphalerons leading to the 
final production of many energetic muons~\cite{Brooijmans:2016lfv}. 
This writeup compares the RFT Monte Carlo predictions to a basic set of QCD observables measured recently
in proton and nuclear collisions at the LHC (Sec.~\ref{sec:2}), and summarizes the results of a recent 
study~\cite{dEnterria:2018kcz} that, for the first time,
interfaced \conex\ with the standard p-p collider MC \pythia~6~\cite{Sjostrand:2006za,Skands:2010ak} to assess the impact of heavy-quark and 
hard jet production on the EAS muon production (Sec.~\ref{sec:3}).

\section{LHC data versus UHECR Monte Carlo generators}
\label{sec:2}

The depth of the maximum of the EAS depends mainly on the characteristics of multiparticle production in the first 
few generations of hadronic interactions in the atmosphere of the incoming UHECR. Among the key observables in hadronic collisions with 
biggest impact on the longitudinal development of EAS~\cite{Ulrich:2010rg} are:
(i) the p-p inelastic cross section  $\sigma_{\rm inel}$ (and its Glauber extension to proton-air interactions, $\sigma_{\rm pAir}$), 
(ii) the density of charged particles at midrapidity per unit rapidity  $\dNdeta$, and
(iii) the fraction of energy carried at forward rapidities. In addition, 
the average transverse momentum of hadrons $\meanpt$ is a sensitive probe of the underlying modeling of pQCD processes.
The level of agreement between the MC predictions for such observables and the first LHC data (mostly p-p collisions
at 7 and 8 TeV) was carefully reviewed in~\cite{dEnterria:2011twh}, where it was found that the 
pre-LHC models overall bracketed the experimental measurements and only small modifications, such as those that 
led to the new \qgsjetII, \eposlhc\ and \sibyll~2.3c releases, were required. A few representative recent experimental results from LHC, 
in particular, pp at $\sqrts = 13$~TeV, {\it not} included in the current Run-1 MC parameter tuning, are compared to the MC predictions below.

\begin{figure*}[htbp!]
\centering
\includegraphics[width=0.39\textwidth,height=6.5cm]{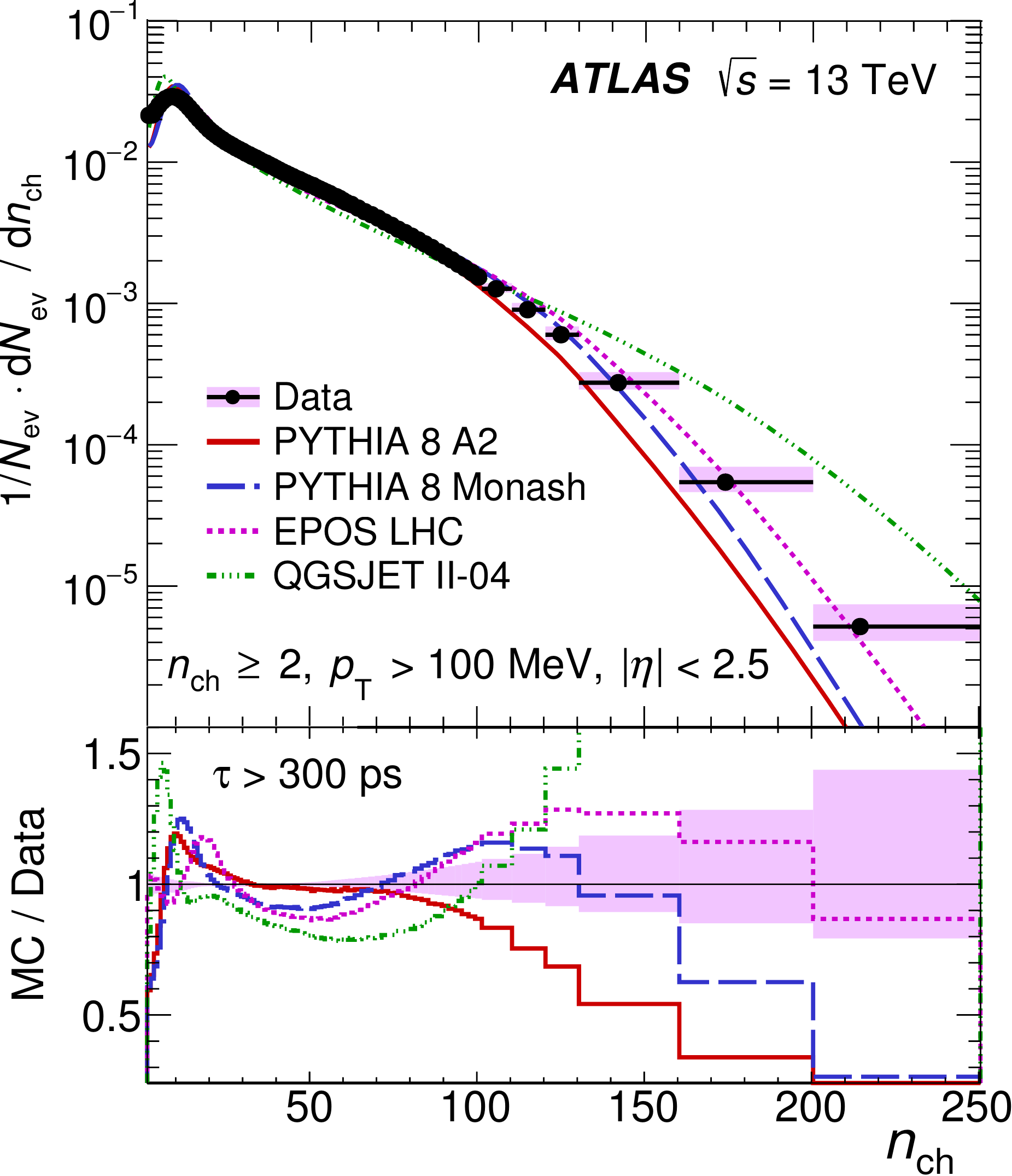}
\includegraphics[width=0.60\textwidth]{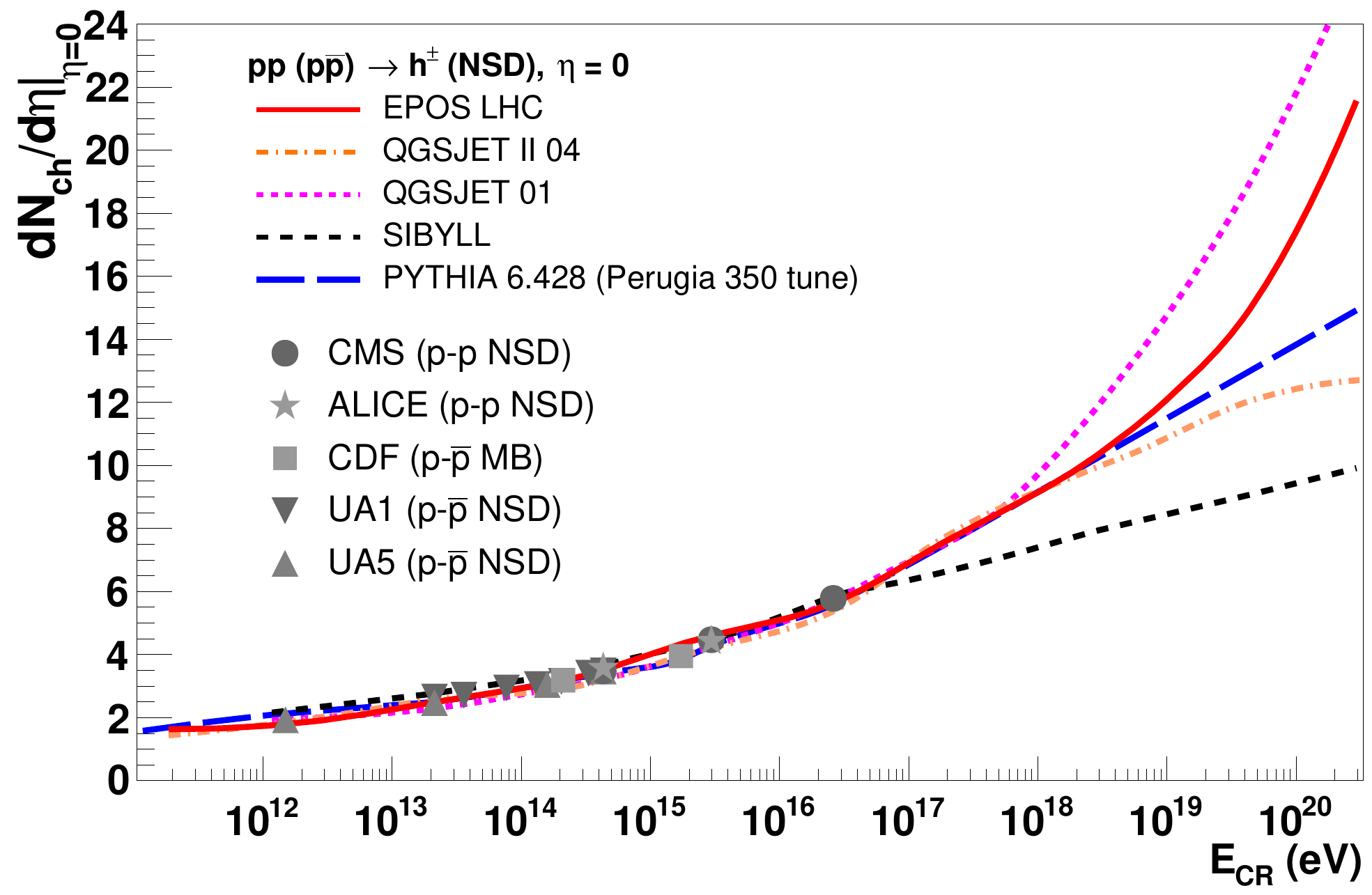}
\caption{Left: Per-event charged-particle multiplicity probability in p-p at $\sqrts = 13$~TeV
measured by ATLAS compared to the predictions of various MC generators~\cite{Aaboud:2016itf}.
Right: Evolution of the charged-particle pseudorapidity density at midrapidity, $\dNdeta$, as a
function of CR energy: Data points (from the compilations~\cite{dEnterria:2011twh,dEnterria:2016oxo})
are compared to the latest MC predictions~\cite{dEnterria:2018kcz}.}
\label{fig:dNchdeta}
\end{figure*}

\begin{figure*}[hbp!]
\centering
\includegraphics[width=0.58\textwidth]{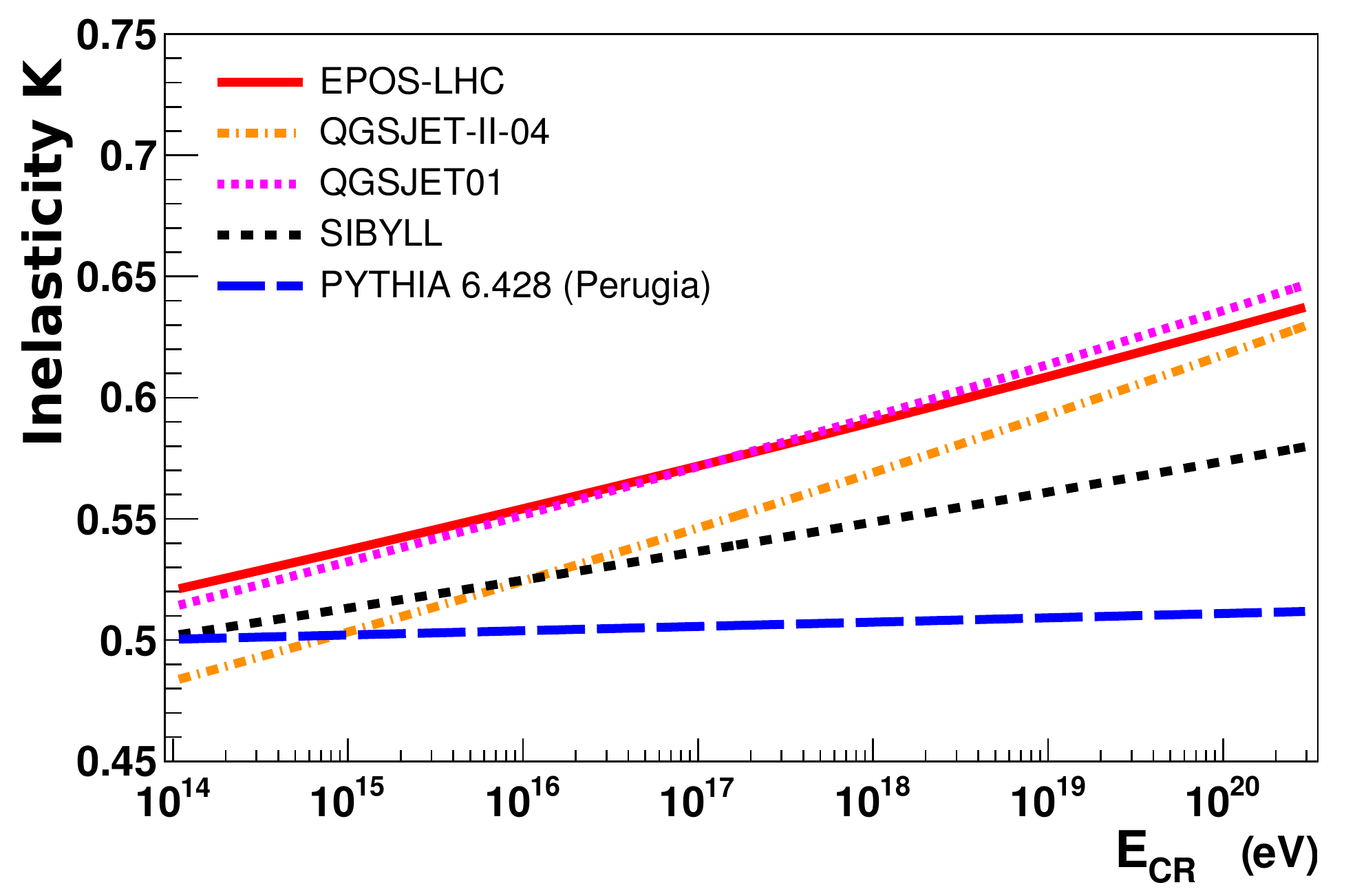}
\includegraphics[width=0.38\textwidth,height=6.4cm]{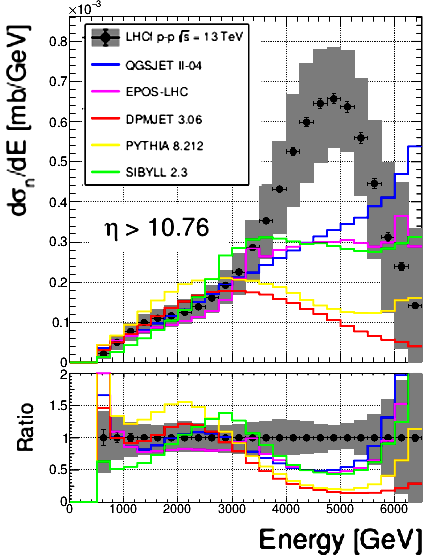}
\caption{Left: MC predictions for the p-p inelasticity as a function of CR energy~\cite{dEnterria:2018kcz}.
Right: Differential neutron cross section measured by the LHCf experiment at $\eta>10.76$ in p-p at $\sqrts$~=~13~TeV
compared to MC predictions~\cite{Adriani:2018ess}.}
\label{fig:fwd}
\end{figure*}

\paragraph{\bf Inelastic p-p cross section}
\label{sec:sigma}

A fundamental quantity of all MC models and key in cosmic-ray physics, as it chiefly determines 
the UHECR penetration in the atmosphere, is the total p-p cross section $\sigma_{\rm tot}$
and its separation into elastic and inelastic (and, in particular, diffractive) components, as well as its 
extension to proton-air interactions. 
Although non-computable from the QCD Lagrangian, the values of $\sigma_{\rm tot,el,inel}$ 
are constrained by fundamental quantum mechanics relations such as the Froissart bound, 
the optical theorem, and the dispersion relations. The pre-LHC models predicted 
$\sigma_{\rm tot}\approx 90$--120~mb at $\sqrts = 14$~TeV depending on whether they preferred to
reproduce the (moderately inconsistent) E710 or CDF measurements at Tevatron~\cite{dEnterria:2011gsx}. 
A large number of $\sigma_{\rm tot,el,inel}$ measurements have been carried out in the last years at the LHC. 
At $\sqrts = 13$~TeV, TOTEM measures 
$\sigma_{\rm tot} = 110.6 \pm 3.4$~mb, with $\sim$72\% inelastic, and $\sim$28\% elastic components~\cite{Antchev:2017dia}.
The MC generators tuned to the experimental results at 7 and 8~TeV reproduce well all existing data 
(Fig.~\ref{fig:sigma_vs_sqrts} left). 
The value of $\sigma_{\rm inel}$ was mostly overestimated by the pre-LHC MC models, leading to an increased 
$\sigma_{\rm pAir}$ cross section, and thereby a smaller $\Xmax$ penetration than predicted now 
by the retuned models. 
In addition, the validity of the geometric Glauber multiple scattering model, used to extrapolate from p-p to p-nucleus 
cross sections~\cite{Auger:2012wt}, has been confirmed by LHC heavy-ion measurements.
 Figure~\ref{fig:sigma_vs_sqrts} (right)
shows the inelastic p-Pb cross section as a function of collision energy with the Glauber prediction (solid black) 
reproducing the $\sigma_{\rm inel,pPb}$ data points measured by CMS at $\sqrtsnn = 5.02$~TeV~\cite{Khachatryan:2015zaa}.

\paragraph{\bf Central particle multiplicity}
\label{sec:mult}

Inclusive hadron production in high-energy proton and nuclear collisions peaks at central rapidities and
receives contributions from ``soft'' and ``hard'' interactions between the partonic constituents of the colliding hadrons. 
Soft (respectively hard) processes involve mainly $t$-channel partons of virtualities $Q^2$ typically 
below (resp. above) a scale $\Qsat^2$ of a few~GeV$^2$. At LHC energies, $\Qsat\approx 2$~GeV in p-p collisions, and 
MC models predict that $\sim$70\% of the final hadrons at midrapidity issue from the fragmentation 
of multiply scattered gluons sharing a low fraction of the colliding hadron momenta,
$x = \rm p_{\rm parton}/p_{\rm had}\approx p_{\rm T,had}/\sqrtsnn\lesssim 10^{-4}$. 
At such low-$x$ and moderate-$Q^2$ values, the parton densities are subject to non-linear
QCD (gluon saturation) effects, in particular in the dense nuclear case~\cite{Albacete:2014fwa}. 
Particle multiplicities are thus particularly sensitive to the modeling of the gluon density at low $x$ and of multiparton interactions (MPI).
The models tuned to LHC Run-1 data overall bracket the latest p-p and p-A distributions measured 
(Fig.~\ref{fig:dNchdeta} left), although improvements are still needed in the tails of the 
multiplicity distributions either at low (diffraction dominated) and/or at high (large number 
of MPI) values. The evolution of the charged hadron pseudorapidity density at 
$\eta = 0$ as a function of CR energy is shown in Fig.~\ref{fig:dNchdeta} (right). All models retuned
to LHC Run-1 data reproduce well the value $\dNdeta \approx 5.5$ measured in p-p collisions at 13~TeV.
The different MC predictions, however, start to rapidly diverge above LHC energies and, at the highest
CR energies of 10$^{20}$~eV, vary within $\dNdeta \approx 10$--25.

\begin{figure*}[htbp!]
\centering
\includegraphics[width=0.60\textwidth]{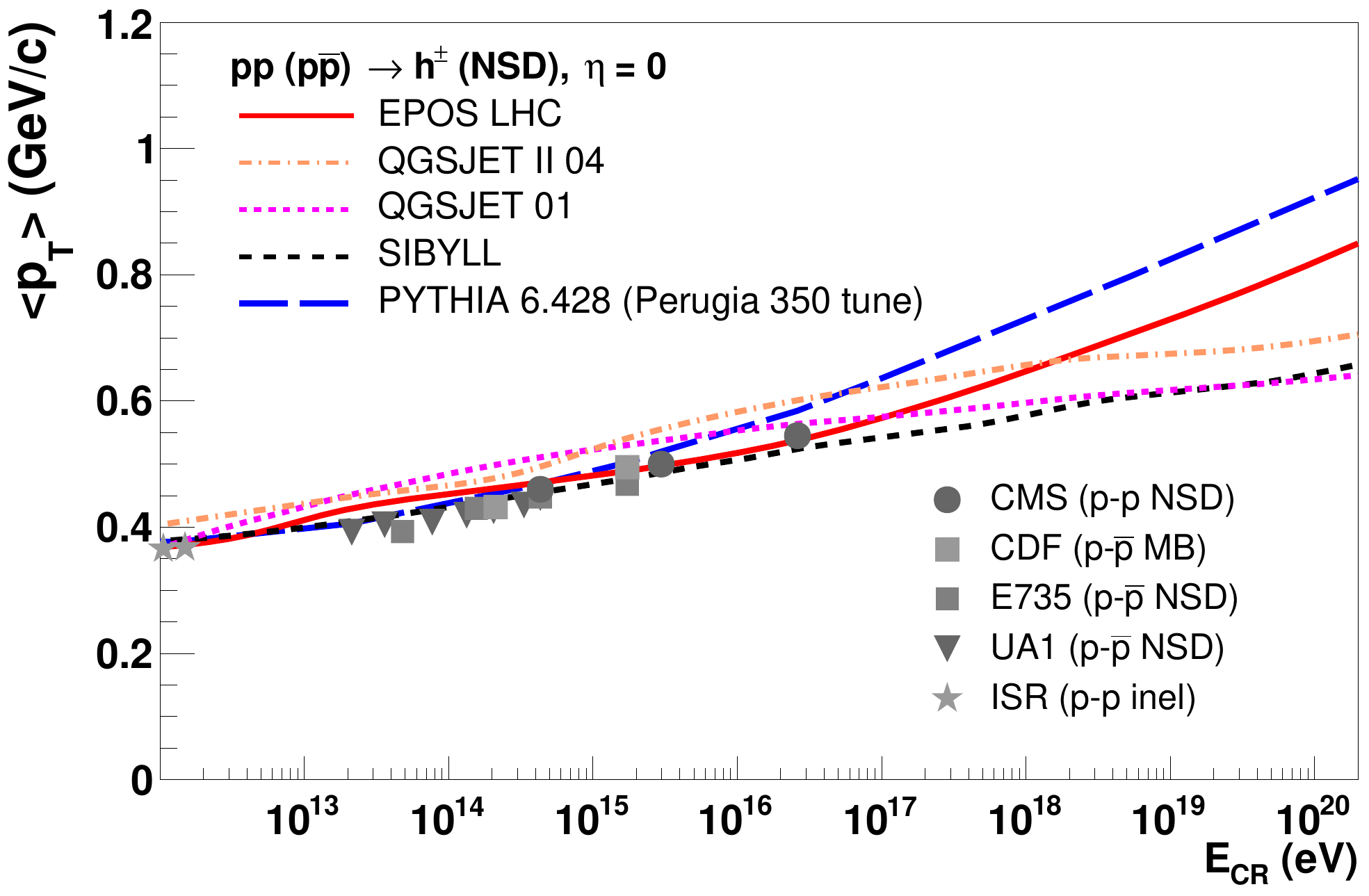}
\includegraphics[width=0.39\textwidth,height=6.5cm]{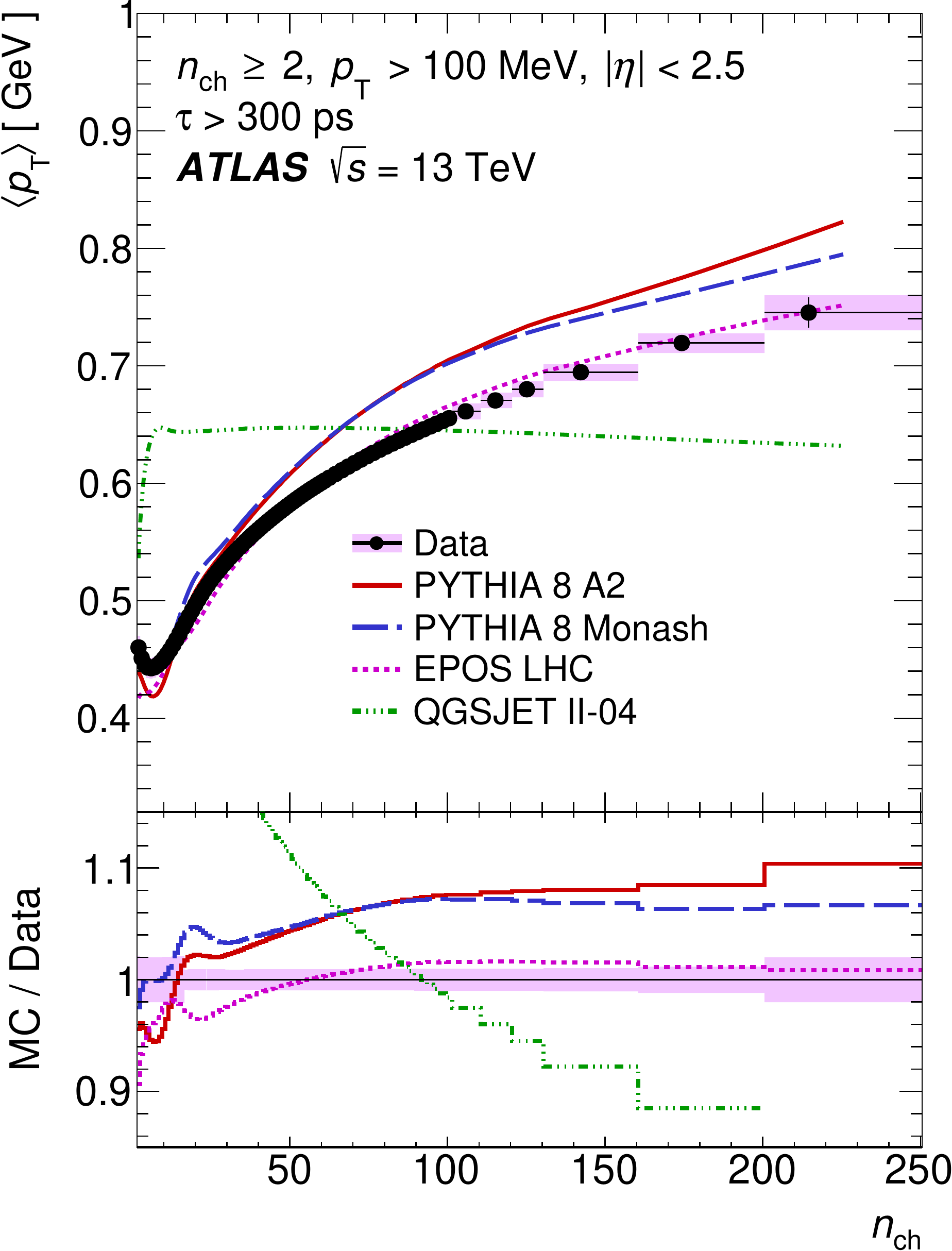}
\caption{Left: Data--model evolution of $\meanpt$ at midrapidity as a function of primary CR energy~\cite{dEnterria:2018kcz}.
Right: Data--model comparison of the mean transverse momentum vs.\,charged-particle multiplicity in p-p collisions at 
$\sqrts$~=~13~TeV~\cite{Aaboud:2016itf}.}
\label{fig:pT}
\end{figure*}

\paragraph{\bf Forward particle production}
\label{sec:fwd}

The fraction of the primary CR energy transferred to secondary particles after removing the most energetic 
``leading'' hadron emitted at very forward rapidities, called inelasticity $K = 1 - E_{\rm lead}/\ECR$,
has an important impact on the EAS development. The model predictions for the CR-energy dependence of the inelasticity 
are shown in Fig.~\ref{fig:fwd} (left): \eposlhc, \qgsjetII, and \qgsjet~01\ feature a $\sim$25\% increase between 10$^{14}$~eV 
and 10$^{20}$~eV, at variance with the almost flat behaviour observed for \pythia~6. Since $\Xmax$ and its average 
fluctuation $\smax$ are mostly driven by the values of $\sigma_{\rm inel}$ and inelasticity, 
proton-EAS simulated with RFT models feature smaller penetration in the atmosphere than those predicted by \pythia~6.
Of particular interest is the forward particle (in particular baryon~\cite{Pierog:2006qv}) production because of 
its key role feeding the muonic component~\cite{Ostapchenko:2016ytp}.
The LHCf experiment has recently measured neutrons at very forward rapidities ($\eta>10.76$) in p-p collisions at 13 TeV~\cite{Adriani:2018ess}
finding that none of the current MC models can reproduce their total nor differential yields (Fig.~\ref{fig:fwd} right).

\paragraph{\bf Transverse momentum spectra}
\label{sec:}

In hadronic collisions, the partonic cross sections peak around the saturation scale, $\meanpt$ $\approx \Qsat$, 
where multiple minijets (of a few GeV) are produced that subsequently fragment into lower-$\pT$ hadrons. This regime
dominated by gluon saturation is modeled in various effective ways in the MC generators. The \pythia\ MC has an 
energy-dependent $\pT$ cutoff separating hard from soft scatterings that mimics the (slow) power-law evolution of 
$\Qsat$~\cite{Albacete:2014fwa}, whereas \sibyll\ uses a logarithmically-running $\Qsat(s)$ scale, and \epos\ and 
\qgsjet\ have a fixed $\pT$ cutoff and low-$x$ saturation is implemented through corrections to the multi-Pomeron dynamics. 
These different behaviours are seen in the energy evolution of the predicted average $\pT$ in Fig.~\ref{fig:pT} (left): 
\pythia~6 shows a faster $\meanpt$ increase  than the rest of models, approaching $\meanpt\approx 1$~GeV at the highest energies. 
All RFT models predict a small rise of $\meanpt$ plateauing at $\sim$0.6~GeV, 
except \eposlhc\ that includes final-state collective parton flow and thereby $\meanpt$ increases with 
multiplicity (and thus $\sqrts$) as seen in Fig.~\ref{fig:pT} (right). Whereas \eposlhc\ and \pythia~6 reproduce the $N_{\rm ch}$-
dependence of $\meanpt$ (through different final-state mechanisms: collective flow and colour reconnection), \qgsjetII\ clearly
misses the data. If one included more pQCD activity in the RFT models, at the expense of smaller final-state effects in the
case of \eposlhc, and/or a nuclear-size dependent $\Qsat$ value in the initial state (as expected on general grounds)~\cite{Albacete:2014fwa}, 
$\meanpt$ would increase and approach the \pythia\ p-p predictions.

\begin{figure*}[htbp!]
\centering
\includegraphics[width=0.49\textwidth]{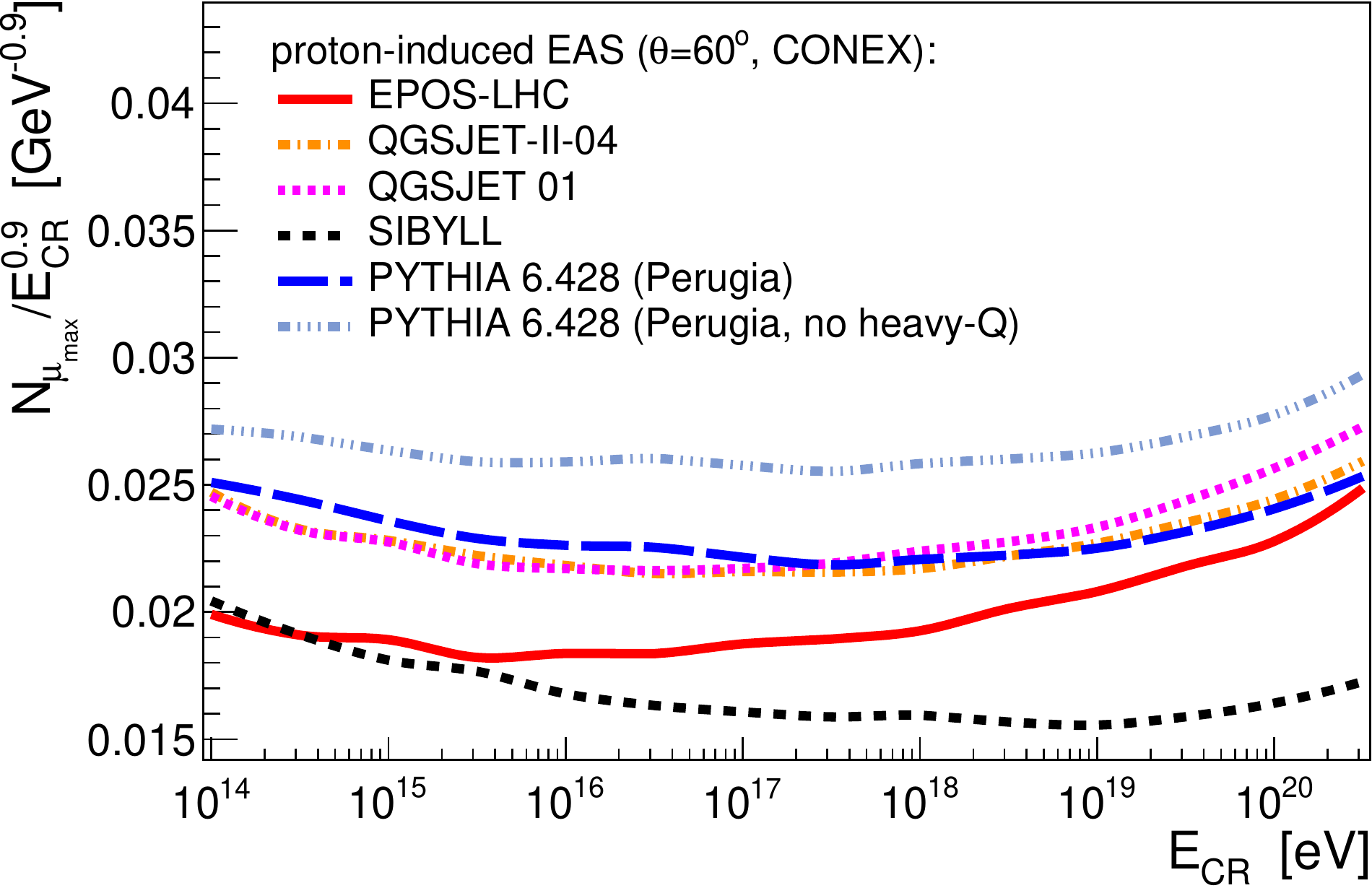}
\includegraphics[width=0.49\textwidth]{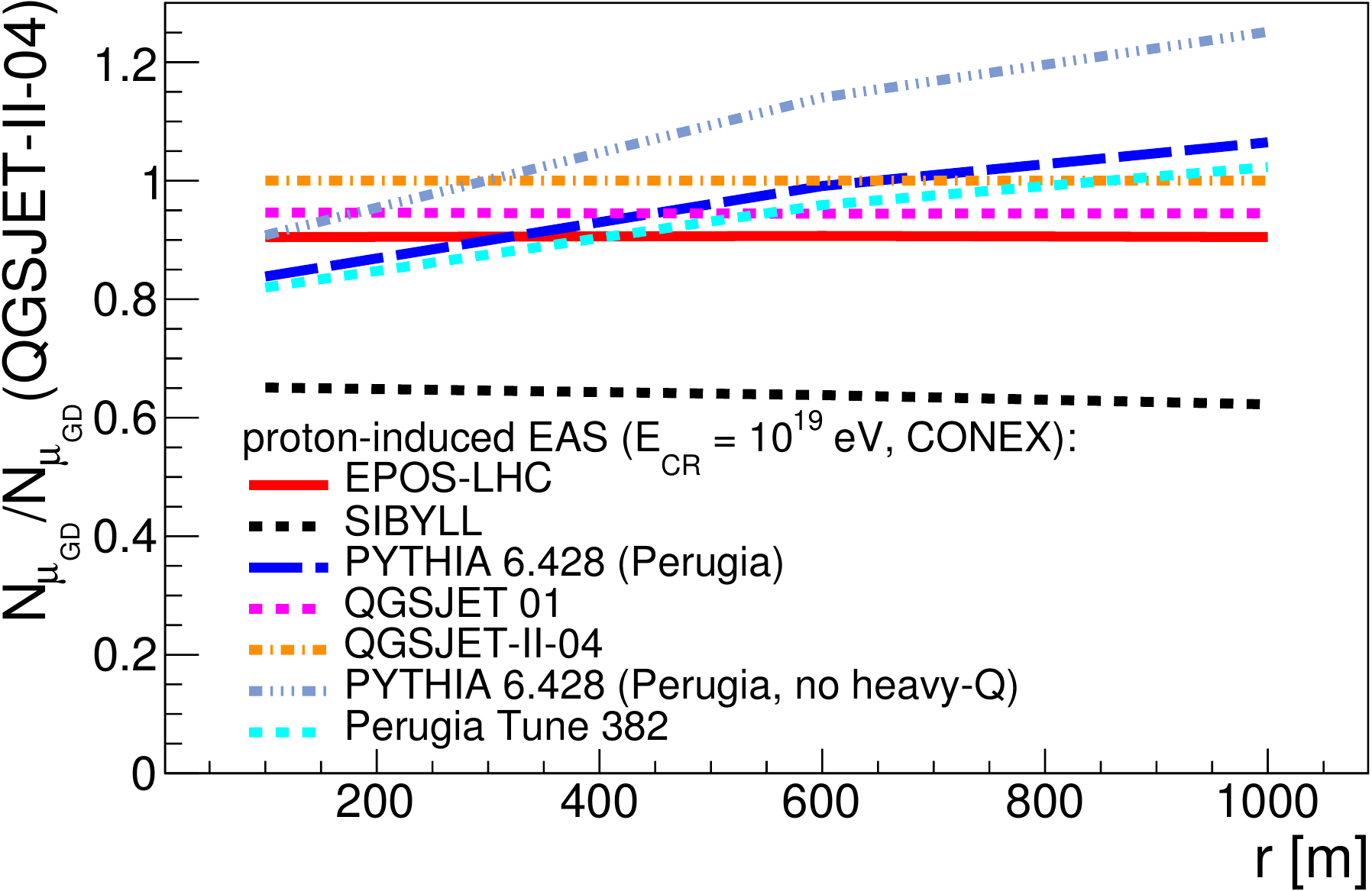}
\caption{Left: Number of muons (normalized by $\ECR^{0.9}$) at shower maximum for inclined proton-induced showers 
as a function of CR energy predicted by different MC generators~\cite{dEnterria:2018kcz}.
Right: Number of muons at sea-level as a function of radial distance to the shower axis for proton-induced 
EAS of $\ECR = 10^{19}$~eV predicted by different MC generators {\it over} the \qgsjetII\ value~\cite{dEnterria:2018kcz}.
\label{fig:muons}}
\end{figure*}

\section{Muon anomalies: Data vs. MC simulations}
\label{sec:3}

Starting at $\ECR\approx 10^{17}$~eV and with increasing primary CR energy, various measured characteristics of muon 
production (accessible in inclined EAS) are difficult to reconcile with the MC models: (i) the observed number 
of muons is underestimated by $\sim$30\% (and even larger values far away from the shower axis), 
and (ii) the measured muon penetration depth is overestimated (indicating inconsistent primary 
composition as derived from the total and $\mu$ shower components)~\cite{Dembinski18}. 
In~\cite{dEnterria:2018kcz}, we studied the production of decay muons from hard QCD jets and heavy-quarks, 
absent in the RFT models, using \pythia~6 to generate proton-EAS in a hydrogen atmosphere with the same density as air. 
In general, the \pythia~6 results for the muon densities and energies at sea level are found in between those predicted 
by the RFT MCs, 
although switching-off heavy-quark production leaves more room for charged pion and kaon production, leading to
a 15\% muon increase on the ground (Fig.~\ref{fig:muons} left). Also, whereas \pythia~6 showers feature a bit 
less muons close to the core ($\lesssim$200~m), their yield is 10--30\% larger at 600--1000~m from the 
shower axis (Fig.~\ref{fig:muons} right). These results indicate that the main $\mu$ sources in \pythia~6 
are the decays of light-quark mesons (charged $\pi,\,k$) from minijet fragmentation,
and that heavy-quarks account for a negligible fraction of the inclusive muons. 
Although part of the muon excess observed at large radii can be solved adding harder minijet activity in the
RFT models, for real UHECR collisions on air, and at variance with the results found in the proton-hydrogen setup, 
\eposlhc\ produces more $\mu^\pm$ than \qgsjetII. Thus, additional nuclear (and/or other) effects are still
needed to explain the final production of muons observed in the data~\cite{Pierog:2017nes}.

\section{Summary}
\label{sec:4}

The determination of the identity (mass) of the highest-energy cosmic rays reaching earth relies heavily on MC 
hadronic generators with parameters tuned to collider data.
Representative measurements of basic
QCD observables (inelastic cross sections, hadron multiplicity, forward particle production, and mean hadron
transverse momenta) measured recently in proton and nucleus collisions at the LHC have been compared to the predictions
of cosmic-ray MC models. The agreement is overall good, but improvements are needed in the modeling of 
very forward particle production and of semi-hard multiparton interactions. Solving the current
muon anomalies observed in the UHECR data requires improved multiple minijet (but, seemingly, not heavy-quarks, at
least for proton-induced showers) production plus a better overall control of nuclear effects
before any potential new-physics contribution can be isolated.\\

\noindent {\bf Acknowledgments --} Common work and multiple discussions with Tanguy Pierog are gratefully acknowledged.

%
%
%

\end{document}